# Title: MRAnnotator: multi-Anatomy and many-Sequence MRI segmentation of 44 structures

[1,†]Alexander Zhou, [1,†]Zelong Liu, [1]Andrew Tieu, [1]Nikhil Patel, [2]Sean Sun, [2]Anthony Yang, [2]Peter Choi, [1]Hao-Chih Lee, [1]Mickael Tordjman, [1]Louisa Deyer, [3]Yunhao Mei, [1]Valentin Fauveau, [1]George Soultanidis, [1,2]Bachir Taouli, [2]Mingqian Huang, [2]Amish Doshi, [1,2]Zahi A. Fayad, [4,5]Timothy Deyer, [1,2,6]Xueyan Mei

[1]BioMedical Engineering and Imaging Institute, Icahn School of Medicine at Mount Sinai, New York, NY

[2]Department of Diagnostic, Molecular, and Interventional Radiology, Icahn School of Medicine at Mount Sinai, New York, NY

[3]Rotterdam School of Management, Erasmus University Rotterdam, Rotterdam, Netherlands

[4]East River Medical Imaging, New York, NY

[5]Department of Radiology, Cornell Medicine, New York, NY

[6]Windreich Department of Artificial Intelligence and Human Health, Icahn School of Medicine at Mount Sinai, New York, NY

[†]These authors contributed equally to this work.

Corresponding Author: Xueyan Mei, PhD, BioMedical Engineering and Imaging Institute, Department of Diagnostic, Molecular, and Interventional Radiology, Windreich Department of Artificial Intelligence and Human Health, Icahn School of Medicine at Mount Sinai, 1470 Madison Ave, New York, NY 10029, United States (xueyan.mei@mssm.edu).

# ABSTRACT

## Purpose

To develop a deep learning model for multi-anatomy segmentation of diverse anatomic structures on MRI.

## Materials and Methods

In this retrospective study, 44 structures were annotated using a model-assisted workflow with manual human finalization in two curated datasets: an internal dataset of 1518 MRI sequences (843 patients) from various clinical sites within a health system, and an external dataset of 397 MRI sequences (263 patients) from an independent imaging center for benchmarking. The internal dataset was used to train an nnU-Net model (MRAnnotator), while the external dataset evaluated MRAnnotator's generalizability across significant image acquisition distribution shifts. MRAnnotator was further benchmarked against an nnU-Net model trained on the AMOS dataset and two current multi-anatomy MRI segmentation models, TotalSegmentator MRI (TSM) and MRSegmentator (MRS). Performance throughout was quantified using the Dice score.

## Results

MRAnnotator achieved an overall average Dice score of 0.878 (95% CI: 0.873, 0.884) on the internal dataset test set and 0.875 (95% CI: 0.869, 0.880) on the external dataset benchmark, demonstrating strong generalization ($p = 0.899$). On the AMOS test set, MRAnnotator achieved comparable performance for relevant classes (0.889 [0.866, 0.909]) to an AMOS-trained nnU-Net (0.895 [0.871, 0.915]) ($p = 0.361$) and outperformed TSM (0.822 [0.800, 0.842], $p < 0.001$) and MRS (0.867 [0.844, 0.887], $p < 0.001$). TSM and MRS were also evaluated on the relevant classes from the internal and external datasets and were unable to achieve comparable performance to MRAnnotator.

## Conclusion

MRAnnotator achieves robust and generalizable MRI segmentation across 44 anatomic structures. Future direction will incorporate additional anatomic structures into the datasets and model. Model weights are publicly available on GitHub. The external test set with annotations is available upon request.

# KEYWORDS

MRI; magnetic resonance imaging; segmentation; machine learning; deep learning; dataset

## SUMMARY STATEMENT

MRAnnotator provides accurate segmentation of 44 gross anatomic structures on MRI, supported by a diverse, fully annotated dataset for benchmarking multi-class MRI segmentation.

## KEY RESULTS

1. MRAnnotator was trained on 1518 MRI sequences from 843 patients, achieving robust, generalizable segmentation of 44 anatomical structures.
2. MRAnnotator achieves average Dice scores of 0.878 (95% CI: 0.873, 0.884) and 0.875 (95% CI: 0.869, 0.880) on an internal test set and an external test set from an independent institution, respectively.
3. The external test set is available by request and provides a unique and diversified dataset for MRI multi-anatomy segmentation benchmarking.

# MAIN TEXT

## Introduction

MRI has become an essential diagnostic tool due to its ability to provide high-resolution, high contrast 3D medical imaging, with vast amounts of imaging data and associated medical information generated annually. Recent advances in machine learning and computational power have spurred significant research interest in applying deep learning techniques to MRI and other modalities to enhance radiological workflows (1,2). Within these workflows, medical image segmentation is particularly crucial for both clinical and research applications, including disease quantification, abnormality detection, therapy planning, dataset curation, and artificial intelligence model development (3,4). The creation of automated segmentation models thus has the potential to reduce radiological workloads, improve diagnostic accuracy, enhance radiologist capabilities, and accelerate machine learning research in radiology (5).

A primary bottleneck in developing these models is the limited availability of annotated medical imaging data for model training, arising from the need for annotations by trained medical experts, privacy regulations, scarcity of particular pathologies, and resource intensity (6,7). An emerging approach to alleviate this bottleneck in medical imaging is the development of foundation models----AI models trained on massive, diverse datasets that can be adapted to new tasks with limited or no additional data (8). However, this approach is still limited in medical imaging due to domain-specific complexities and limited availability of large-scale, diverse medical datasets. As a result, most current medical imaging models are task-specific, designed to handle narrowly defined tasks (9). Another significant challenge in model development is the degradation of model performance when encountering unseen external data with distinct image acquisition characteristics, a limitation often referred to as the “scanner effect” (10,11).

Recently, Kirillov et al. proposed the concept of “Segment Anything,” demonstrating that models trained on a large-scale segmentation dataset can be generalized across a wide range of downstream segmentation tasks (12). Developing MRI-specific segmentation models that support a broad range of labels could optimize computational resources for multiple downstream applications. Furthermore, creating robust, generalized segmentation models would enable the integration of complementary datasets, facilitating the creation of larger, more unified datasets and models.

To address this need, we aim to develop a modality-specific model capable of multi-class 3D anatomic segmentation on MRI imaging data with robust performance across various anatomic regions (MRAnnotator). Following the approach established by Wasserthal et al. with TotalSegmentator for CT (13), we trained MRAnnotator on a large, diverse internal MRI dataset and evaluated its generalizability using an external dataset from a separate medical center. We further demonstrate the utility of MRAnnotator by leveraging its pretrained weights and benchmarking on a well-known public dataset against existing state-of-the-art multi-class MRI segmentation models.

## Materials and Methods

This retrospective study was performed in accordance with relevant local and national regulations and approved by STUDY-24-00008 at the Icahn School of Medicine at Mount Sinai. The requirement to obtain individual informed consent was waived.

### Dataset Curation

Two datasets were curated and annotated for this study. The first dataset consisted of studies from our academic health system for the training, validation, and testing of MRAnnotator ("*internal dataset*"). The second dataset consisted of studies from an independent outpatient radiology facility to test model generalizability across scanner variations ("*external dataset*").

***Internal dataset.*** To create a robust and diverse dataset for model development, 678 studies were selected from the internal PACS of the Mount Sinai Health System across six clinical sites throughout New York, NY, between 2010 and 2023. Studies were manually reviewed by a third-year radiology resident to exclude notable artifacts, major pathologies, and lack of target sequences. An additional 171 lumbar spine studies from a prior research study were included, totaling 849 studies from 843 unique patients. Multiple sequences were selected from each study for a total of 1518 sequences. Additional details on study protocols and sequences are shown in **Table 1**. Image preprocessing was handled by the nnU-Net pipeline (3). The final dataset was divided by patient into training (865 sequences), validation (217 sequences), and a holdout test sets (436 sequences) (**Figure 1A**).

***External dataset.*** 291 studies were selected from the PACS of East River Medical Imaging/RadImageNet LLC, a separate clinical imaging center, between 2006-2022, for external testing and evaluation of model generalizability. Studies were reviewed by a senior radiologist to exclude notable artifacts, major pathologies, insufficient resolution, and different FOVs. The same sequence selection process yielded a total of 397 MRI sequences. Although similar sequences were selected, imaging was acquired on different machines with different parameters between the two sites; additional details are shown in **Table 1**. Image preprocessing was similarly handled by the nnU-Net pipeline. This entire dataset was designated as a holdout test set for external validation (**Figure 1A**).

## Dataset Annotation

44 anatomic structures in 8 anatomic regions were annotated (**Figure 2, Table 1**). Segmentation was conducted by 3 medical students (annotators) under supervision of 2 senior radiologists each with 10+ years of clinical experience, who approved and corrected all annotations as needed. The Discovery Viewer platform was used for manual segmentation, as well as the generation and manual correction of intermediate model-assisted annotations (14).

To speed up the annotation process, we adopted an iterative model-assisted workflow similar to Wasserthal et al. for all non-vertebral anatomic structures (13) (**Figure 1B**). 10 initial sequences were manually segmented by annotators and used to train a preliminary nnU-Net model (3). This was used to generate intermediate segmentations for an additional 10 sequences, which were then corrected by annotators. The intermediate annotation model was then retrained on all 20 sequences. This correction and retraining process was repeated with 40 and 60 total sequences, and then used to generate intermediate segmentations for all 878 non-vertebral sequences. For vertebral anatomic structures (C3-C7, T1-T12, L1-L5), annotators used MedSAM (15) to generate intermediate segmentations from bounding boxes for all 640 vertebral sequences. MedSAM was used for this task due to its ability to segment the relatively well-defined and higher-contrast vertebral bodies.

This workflow was used to annotate both the internal and external datasets. Each dataset was annotated independently (i.e., completely separate intermediate annotation models/workflows) to avoid introducing potential bias. All 1518 internal sequences and all 397 external sequences were ultimately manually reviewed and corrected by annotators, with final review and correction by supervisors. These finalized annotations served as the ground truth for model training and internal/external evaluation. Examples of finalized annotations are shown in **Figure 2** and **Figure 3**.

## Model Training

MRAnnotator uses a 3D nnU-Net architecture, which self-configures parameters based on dataset properties and hardware conditions (3), achieving robust performance across various medical imaging applications (13,16,17). The model was trained on the finalized internal dataset, independent of intermediate annotation models. For the configuration of the training strategy, we manually disabled mirroring for data augmentation to help the model learn the laterality of 3D volumes (e.g., differentiating right vs. left structures).

For efficient training, we split the 44 classes into four sub-models targeting specific anatomic studies (abdomen; pelvis + prostate; shoulder + knee; cervical + thoracic + lumbar spines). This segmentation strategy reduced training time, achieving convergence in ~1000 epochs compared to ~4000 epochs with improved performance (supplementary **S2**). Model weights with the highest Dice score on the validation dataset were saved as the finalized model for evaluation. The training was done on an NVIDIA DGX A100 80 GB.

## Model Evaluation

***Internal and external dataset performance.***

Model performance was evaluated on the holdout test set from the internal dataset, unseen by the model during training and hyperparameter tuning. Model performance on all segmentation classes was also evaluated on the external dataset, which was annotated with the same anatomic classes. The entire external dataset was used as a holdout test set, and the model did not see it during training.

***Model benchmarking.*** To further assess model performance and robustness, we evaluated MRAnnotator on the AMOS abdominal MRI dataset (19). We compared MRAnnotator against an nnU-Net model trained on AMOS, measuring average Dice score on relevant structures.

Additionally, we compared MRAnnotator with two state-of-the-art multi-anatomy MRI segmentation models: TotalSegmentator MRI (TSM) (20) and MRSegmentator (MRS) (21). To demonstrate the utility of MRAnnotator alongside these models, we use the publicly available pretrained weights of TSM and MRS to benchmark relative performance on AMOS and our external dataset. We benchmark on the AMOS dataset because it is a well-established dataset used for model evaluation on abdominal MRI segmentation. We also benchmark on our external dataset, as it represents a robust and diversified dataset with many anatomical structure classes, making it well-suited for evaluation of these multi-anatomy MRI segmentation models.

The AMOS nnU-Net model was trained and evaluated with 5-fold cross validation on the AMOS dataset, while MRAnnotator, TSM, and MRS were directly evaluated on the full AMOS dataset with zero-shot evaluation.

***Statistical Analysis***. The Dice similarity coefficient (“Dice score”), which measures spatial overlap accuracy, was used to evaluate model-predicted segmentations compared to the corresponding human-finalized ground truth segmentations (18). A Dice score of 0 indicates no overlap, while 1 indicates perfect overlap.

We evaluated the normal distribution of Dice scores for each anatomic structure via the Shapiro-Wilk test with a p value threshold of 0.05. If the data were normally distributed, we calculated the 95% confidence interval (CI) for the Dice score using the central limit theorem. If not, we used a bootstrapping method involving 1,000 resamples to calculate the 95% CI.

A non-inferiority test compared performance between internal and external test sets, with a delta margin of 1% of the internal Dice score mean value; a p value $< 0.05$ rejects the null hypothesis that MRAnnotator demonstrates non-inferior performance on external datasets compared to the internal dataset. Paired t-test were performed for statistical significance.

### Data Availability

Model weights are publicly available on GitHub (github.com/lzl199704/MRAnnotator) to facilitate out-of-the-box use or other downstream applications. The external dataset with annotation is available for public use by request, under a data usage agreement. By making this unique dataset available, we hope to provide a diversified benchmark for model evaluation across multiple anatomic structures, anatomic regions, and MRI sequences. Due to our institutional private health information policies, we are unable to share the internal dataset.

## Results

### Dataset Characteristics for Training and Generalizability

**Figure 4** shows the distribution of patient demographic and image acquisition characteristics across both internal and external datasets. The internal dataset, used for model development, has a mean age of 50.3 (IQR 34-65), with 438 females and 411 males. It demonstrates broad diversity in patient and image acquisition, with significant representation from multiple MRI manufacturers and models. The external dataset has a mean age of 58.5 (IQR 48-71) with 127 females and 164 males. The external dataset represents a significant shift in demographics and image acquisition characteristics, enabling the evaluation of model generalizability to new sites with a distribution shift.

### Segmentation Evaluation

***Internal and external dataset performance.***

MRAnnotator achieved an overall average Dice score of 0.878 (95% CI: 0.873, 0.884) on the internal dataset holdout test set and an overall average Dice score of 0.875 (95% CI: 0.869, 0.880) on the external dataset, demonstrating strong generalizability ($p = 0.899$).

Model performance and generalizability vary by anatomic study. Average results broken down by anatomic study protocol are summarized in **Table 2**, while performance on each individual anatomic structure is shown in **Figure 5** and summarized in supplementary **S1**. Example failure cases are shown in **Figure 6**. Notably, MRAnnotator achieves acceptable performance on all anatomies; it also achieves non-inferiority on the external dataset in all anatomic studies, aside from pelvis and prostate. This suggests that MRAnnotator generalizes well to external data with independent acquisition protocols, demonstrating its robustness to scanner-related domain shifts.

Additionally, we evaluate our model's performance across different field strengths and manufacturers, summarized in **Table 3**. Notably, performance is relatively consistent at field strengths above 1.5 T but is reduced at a lower field strength of 0.55 T. This is likely due to a combination of lower resolution and fewer low-field training data. Lastly, performance is consistent across well-represented manufacturers but is notably reduced on an unseen manufacturer (Hitachi) in the external dataset. We note that all the 0.3 T studies were acquired on Hitachi machines, and vice versa. These studies were all acquired on open MRI models, which have lower image quality and likely explains the reduced model performance on 0.3 T / Hitachi scans.

***Model benchmarking.***

We trained a separate nnU-Net model on the AMOS dataset ("AMOS nnU-Net") and compared its performance against MRAnnotator on the AMOS, internal, and external test datasets (8 common abdominal classes). 5-fold cross validation was used to train and evaluate AMOS nnU-Net performance. MRAnnotator, TSM, and MRS were directly evaluated on the AMOS dataset with zero-shot evaluation.

MRAnnotator achieved comparable performance on the AMOS test set (Dice: 0.889; 95% CI: 0.866, 0.909) compared to AMOS nnU-Net (Dice: 0.895; 95% CI: 0.871, 0.915) ($p = 0.361$). On the internal test set, MRAnnotator achieved significantly higher performance (Dice: 0.915; 95% CI: 0.902, 0.928) compared to AMOS nnU-Net (Dice: 0.805; 95% CI: 0.780, 0.826), as well as on the external dataset (MRAnnotator Dice: 0.912; 95% CI: 0.898, 0.923 versus AMOS nnU-Net Dice: 0.879; 95% CI: 0.858, 0.897) ($p < 0.001$ and $p < 0.001$, respectively). These results demonstrate improved performance of MRAnnotator on an independent benchmark dataset, further validating its generalizability.

We further evaluated MRAnnotator against TSM and MRS. MRAnnotator achieved higher performance than TSM on both the external abdominal dataset (Dice: 0.822; 95% CI: 0.806, 0.837; $p < 0.001$) and AMOS test set (Dice: 0.822; 95% CI: 0.800, 0.842; $p < 0.001$). Similarly, MRAnnotator outperformed MRS on the external abdominal dataset (Dice: 0.869; 95% CI: 0.854, 0.882; $p < 0.001$) and AMOS test set (Dice: 0.867; 95% CI: 0.844, 0.887; $p < 0.001$). Similarly, MRAnnotator achieved higher performance on the abdominal internal test set compared to TSM and MRS, which was expected, as this was the dataset used to develop MRAnnotator. These results are summarized in **Table 4**. This improvement in relative performance likely reflects the larger training dataset of MRAnnotator, as well as differences in the underlying characteristics of the curated training data for each model.

Lastly, we benchmarked TSM and MRS on the full external dataset and compared against MRAnnotator's performance in order to assess general multi-anatomy performance. We only evaluated performance of TSM and MRS on structures that were labeled in both their respective datasets and the external dataset. Performance is summarized in **Table 5** by anatomic study, with structure-specific performance in supplementary **S3**. MRAnnotator achieved superior performance on relevant anatomic structures, though it is worth noting that TSM and MRS included additional structures not labeled in our internal and external datasets that MRAnnotator would not be able to segment.

## Discussion

In this study, we trained a robust model, MRAnnotator, on a diversified dataset of 1518 MRI sequences that can segment 44 anatomic structures across various anatomic regions. Our model achieved high accuracy across all classes on both internal and external test data, demonstrating strong generalizability. We further demonstrated generalizability of our model on an independent, well-established dataset, underscoring its potential for various downstream tasks. Additionally, MRAnnotator outperformed two leading multi-anatomy MRI segmentation models in relevant structure segmentation.

Despite the availability of MRI datasets of varying size for brain (22–24), abdomen (19,25), pelvis (26), and cardiac (27), among others (28), few studies have developed MRI segmentation models capable of handling multiple structures across varied anatomic regions. Multi-class models trained on these datasets have achieved high performance but remain limited by the number of labeled structures and the anatomic focus within an individual dataset. For example, Ding et al recently proposed a novel state-of-the-art self-supervised few-shot segmentation model on the CHAOS MRI dataset, but it was restricted to four abdominal structures (29). Similarly, Kondo and Kasai's residual U-Net model with deep supervision for CT/MRI segmentation of abdominal organs on the AMOS dataset was limited to 15 abdominal structures (30).

These examples highlight the gap in training generalized multi-anatomy MRI segmentation models that can serve as an out-of-the-box tool or starting point for downstream tasks. Towards this goal, our annotated external dataset provides a standardized and diversified many-class, many-anatomy dataset for the development of these models. MRAnnotator further demonstrates the viability of this approach and contributes to the ongoing development of larger models that can incorporate even more anatomies, modalities, and/or tasks. Our model complements similar multi-anatomy MRI segmentation efforts, by offering alternative labeled structures, relative performance, and underlying dataset characteristics, thereby expanding options and robustness for researchers and clinicians to further build upon.

We anticipate several potential applications for MRAnnotator. As an out-of-the-box tool, it can be used in model-assisted annotation workflows for curating larger labeled datasets, as part of a broader model pipeline, or for automated image annotation in research. MRAnnotator can also be adapted for novel applications via techniques such as transfer learning, fine tuning, few-shot learning, and zero-shot learning, or in combination with emerging techniques such as generative, semi-supervised, or self-supervised approaches. Its consistent performance across diverse datasets demonstrates its applicability in realistic clinical settings, accommodating equipment from multiple vendors and major MRI sequences. This versatility makes MRAnnotator valuable for both research and clinical applications, with potential to support routine operations and quantitative analyses essential for longitudinal studies and multicentric trials.

Our work has several limitations. Firstly, we used specific sequences from each patient study, whereas multiple sequences are typically used for diagnosis and segmentation. We were not able to include all sequences for every protocol in this study, which may reduce performance on unseen MR sequences (e.g., undersampled rapid MRI sequences). We also incorporated a single plane for each anatomic study protocol, while comprehensive MRI analyses often utilize multiple sequences and planes. Additionally, although MRAnnotator demonstrated strong generalization to external data due to the robustness of our curated datasets, we directly employed the nnU-Net framework – more novel architectures and techniques could further improve performance, but this was beyond the scope of this project. Lastly, our iterative model-assisted segmentation workflow enabled us to generate labels more efficiently; however, this approach may reduce inter-observer variability, potentially yielding slightly higher Dice scores on average. An inter-observer variability assessment of the ground truth annotations was not conducted.

Future work will focus on several directions. Firstly, we aim to incorporate additional sequences, planes, and anatomic regions, such as chest, cardiac, musculoskeletal, and brain, into our model and datasets. This could enhance performance on existing classes and broaden the utility of MRAnnotator. Secondly, we plan to investigate the utility of our model in the preparation of large-scale annotated datasets for the training of larger-scale models for medical imaging. Lastly, we plan to apply the data and models from this and future work towards the development of larger-scale and more generalized models.

In conclusion, we developed MRAnnotator, an MRI segmentation model that is easy to use, capable of segmenting 44 anatomic structures throughout the body, and generalizable to unseen external imaging center data. Furthermore, we have made our external dataset available as diversified benchmark for MRI multi-anatomy segmentation.

**Disclosure of Funding**

None

**Conflict of Interest**

T.D. is the managing partner of RadImageNet LLC and a paid consultant to GEHC and AirsMedical. X.M. is a paid consultant to RadImageNet LLC.. X.M is a paid consultant to RadImageNet LLC.

**Acknowledgements**

X.M is supported by the Eric Schmidt Foundation.

**Data/Code Availability**

Model weights are publicly available on GitHub (https://www.github.com/lzl199704/MRAnnotator). The external dataset from East River Medical Imaging/RadImageNet LLC with annotation is available for public use by request (https://www.radimagenet.com/), under a data usage agreement. Due to our institutional private health information policies, we are unable to share the internal dataset.

## Figures and Captions

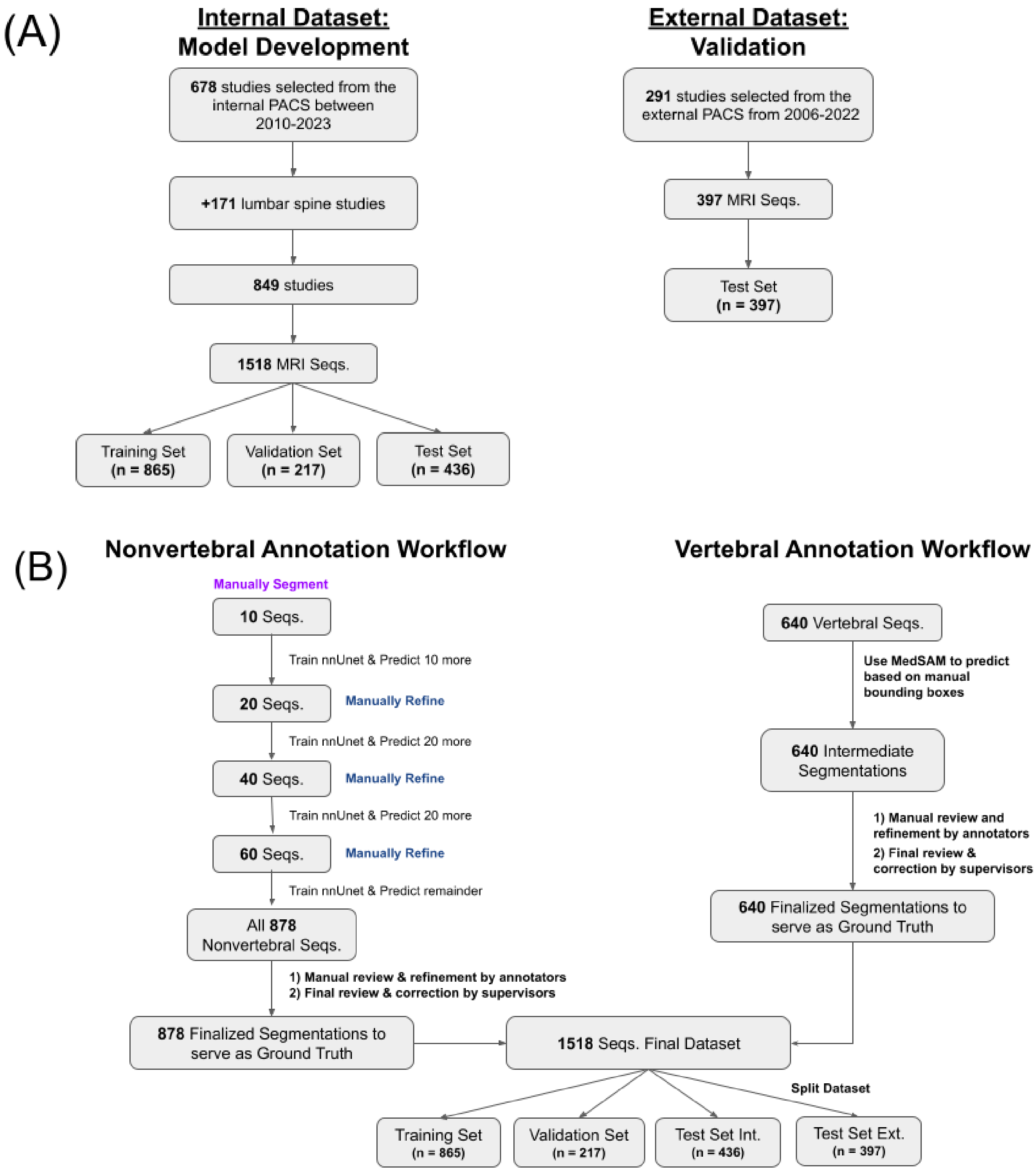

**Figure 1:** (A) Diagram illustrating the patient inclusion criteria and sequences selection process used for the internal dataset's curation and annotation. (B) Diagram depicting the iterative, model-assisted workflow employed for the annotation of anatomical structures. This workflow was independently repeated to annotate the external dataset.

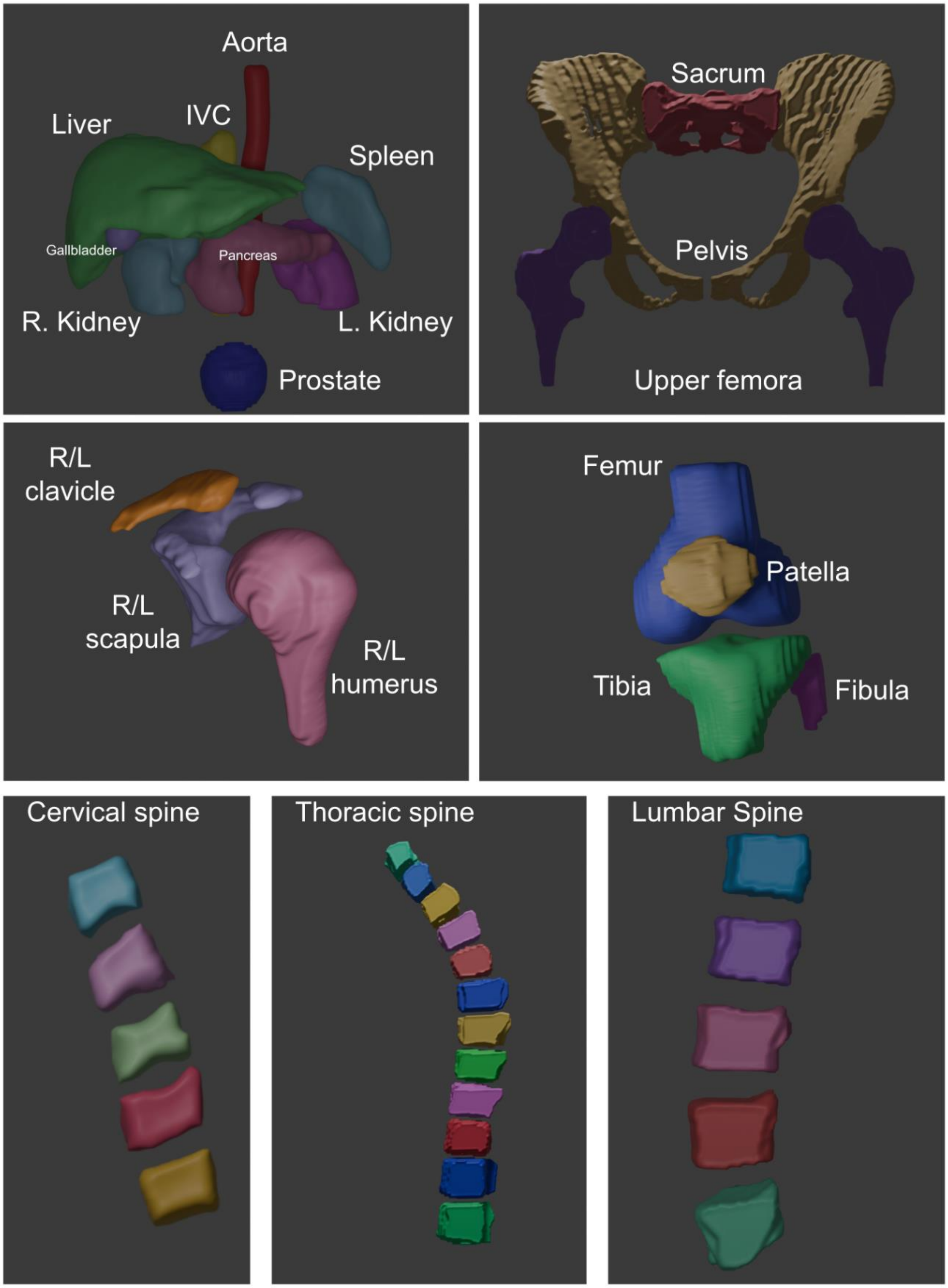

**Figure 2:** Examples of finalized model annotations across the 8 anatomic study protocols, illustrating the 44 labeled classes. For visualization purposes, the prostate was combined with the abdominal structures. These 3D visualizations were generated from real finalized image annotations that were used for subsequent model development and training. R=right L=left

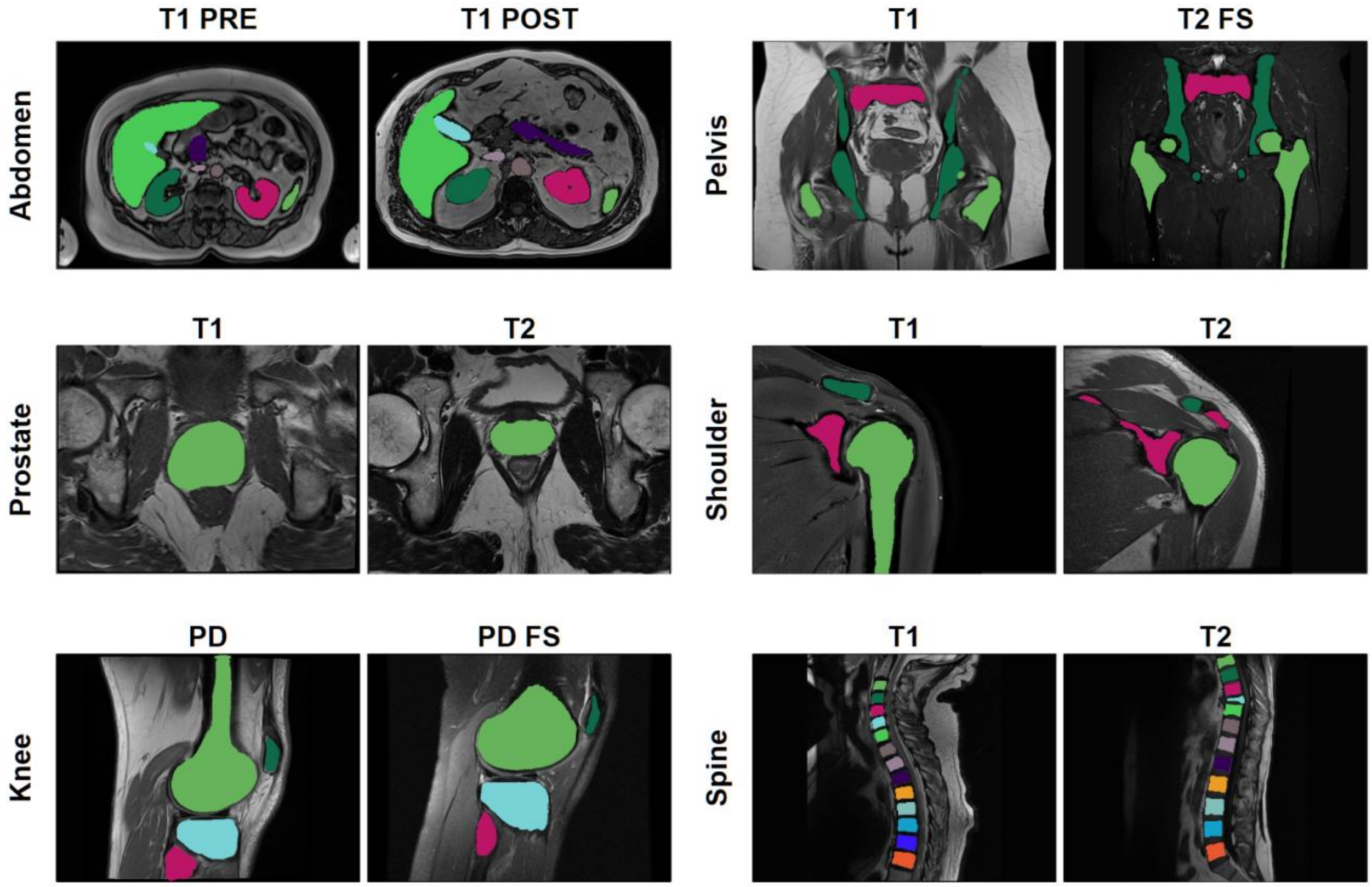

**Figure 3:** Examples of finalized 2D annotations on various anatomic study protocols and sequences. PD: proton density; FS; fat saturation; Pre: pre-contrast; Post: post-contrast

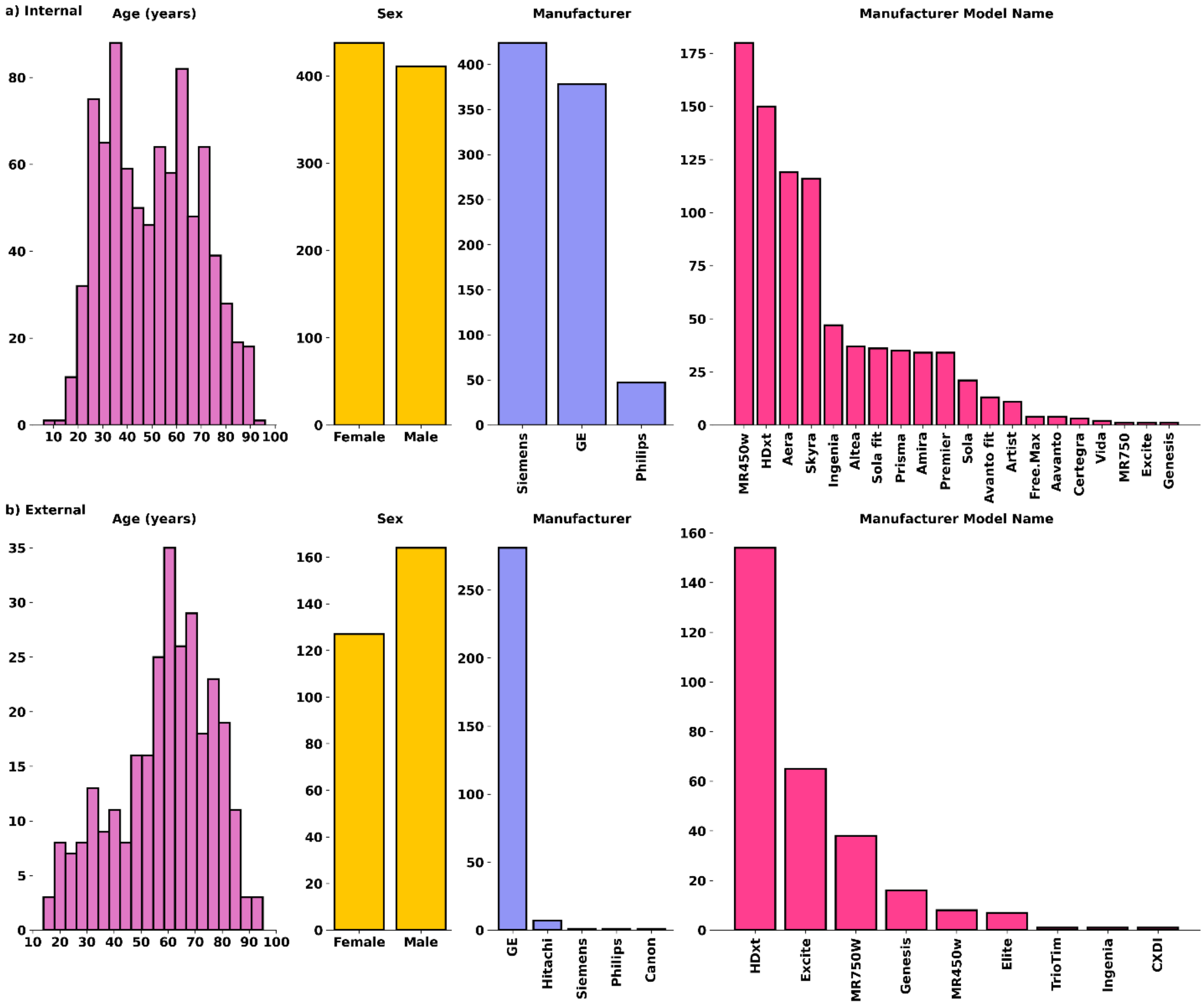

**Figure 4:** Characteristic distributions of **a)** the internal and **b)** the external datasets. These visualizations highlight the substantial diversity present in the internal dataset, which was used for training, as well as the significant distribution shift observed in the external dataset, for generalizability.

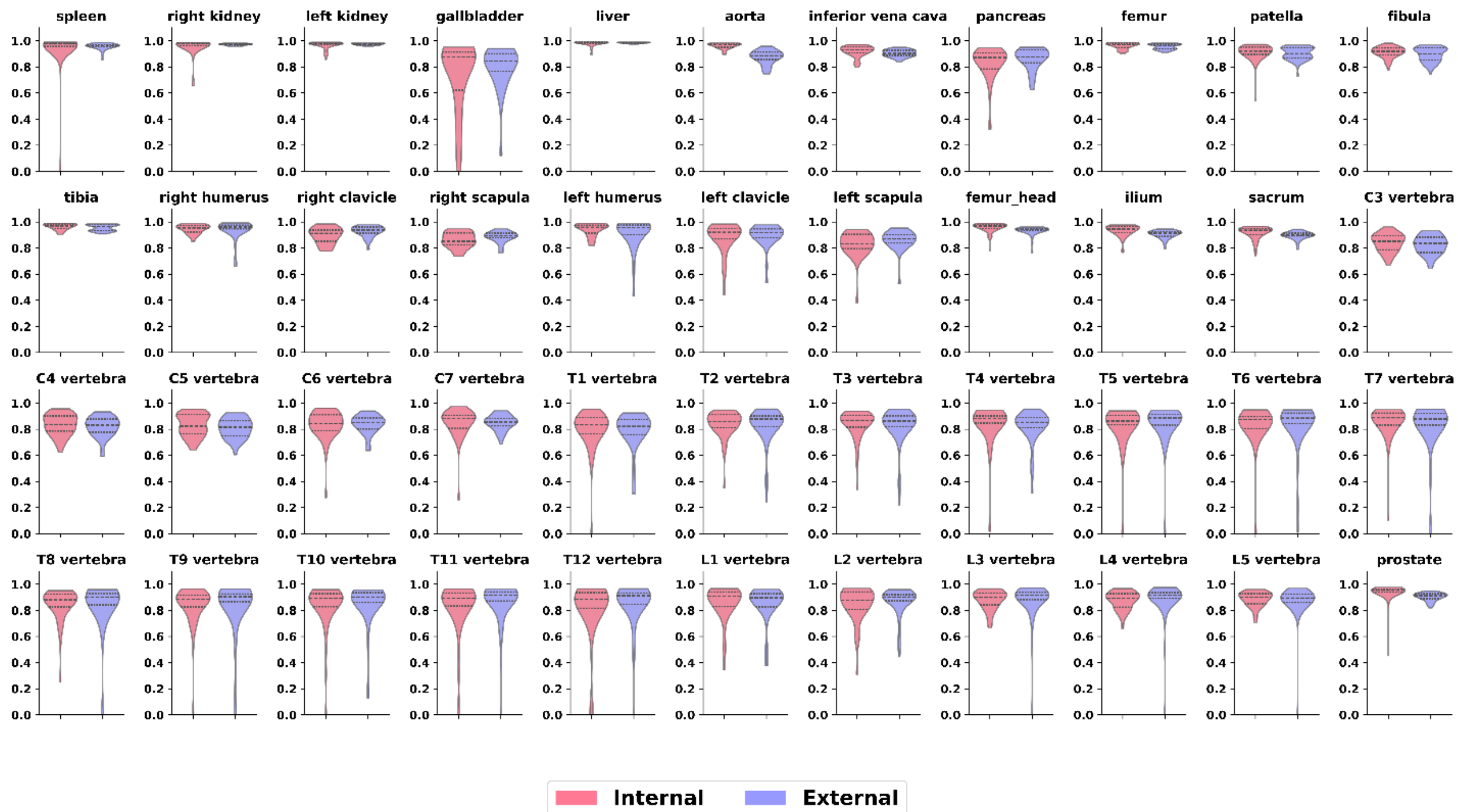

**Figure 5.** Performance of MRAnnotator on all 44 individual classes on the internal and external data test sets, measured by Dice score.

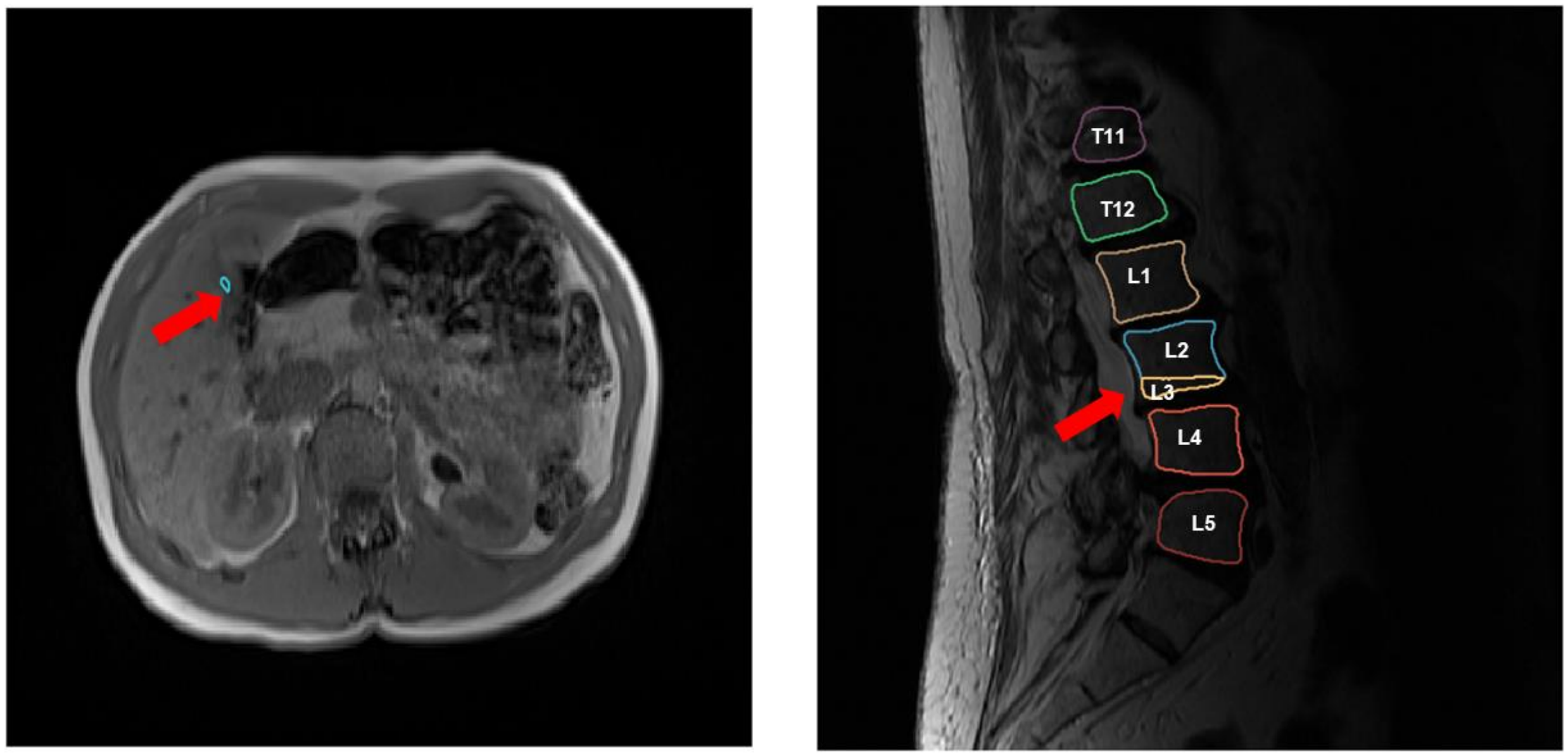

Abdomen L-Spine

**Figure 6.** Representative examples of failed cases. In the abdomen, a failure case observed is the inability to detect the gallbladder structure (false negative), possibly due to an empty gallbladder. In the spine, failure cases included misidentification or incorrect ordering of vertebrae (e.g. swapping L5 with S1) or failed detection of existing vertebrae, which may be attributed to lumbosacral transitional abnormalities and the limitation of our study to sagittal spine views, which offer fewer differentiating features. In our results on the internal test set and external test set, we see almost no complete or near-complete failure cases (i.e. Dice score <0.5) for all other structures.

# Tables

**Table 1:** All 44 anatomic structures segmented and number of included studies.

| Anatomic Study Protocol | Anatomic Structures | Sequences/ Techniques | # Patients, Studies, & Sequences (Internal) | # Patients, Studies, & Sequences (External) |
|---|---|---|---|---|
| Abdomen (axial) | Spleen<br>Right kidney<br>Left kidney<br>Gallbladder<br>Liver<br>Aorta<br>Inferior vena cava<br>Pancreas | T1GRE, In/out of phase, T2, T2 FS, post-contrast, Dixon, LAVA | 89, 89, 174 | 30,30, 30 |
| Pelvis (coronal) | Upper femur<br>Pelvis<br>Sacrum | T1, T2, T2 FS, STIR, PD | 99, 99, 181 | 30, 30, 41 |
| Prostate (axial) | Prostate | T1, T2 | 89, 89, 178 | 59, 60, 60 |
| Shoulder (coronal) | Right humerus<br>Right clavicle<br>Right scapula<br>Left humerus<br>Left clavicle<br>Left scapula | PD, PD FS, T2, T2 FS | 94, 94, 172 | 30, 30, 60 |
| Knee (sagittal) | Femur<br>Patella<br>Fibula<br>Tibia | PD, PD FS, T2, T2 FS | 95, 95, 173 | 30, 30, 58 |
| Cervical spine (sagittal) | Vertebrae C3-C7 | T1, T1 FLAIR, T1 FS, T2, T2 FS, STIR | 111, 111, 181 | 30, 31, 40 |
| Thoracic spine (sagittal) | Vertebrae T1-T12 | T1, T1 FLAIR, T2, T2 FS, STIR | 102, 102, 167 | 31, 32, 44 |
| Lumbar spine (sagittal) | Vertebrae L1-L5 | T1, T1 FLAIR, T1 FS, T2, T2 FS, STIR | 171[a], 171[a], 292[a] | 30, 48, 64 |
| **Total** | **44 structures** | **11** | **843[b], 849, 1518** | **263[c], 291, 397** |

[a]These studies were part of a curated cohort for another project, and all patients in this cohort presented with compression fractures.
[b]six patients appeared in multiple categories and one study contained both cervical/thoracic and lumbar spine sequences.
[c]Seven patients appeared in multiple categories.
Abbreviations: GRE, gradient recalled echo; FS: fat suppressed; STIR: short tau inversion recovery; PD: proton density; LAVA: liver acceleration and volume acquisition; FLAIR: fluid attenuation inversion recovery.

**Table 2:** MRAnnotator model performance on internal and external test datasets, measured by Dice score.

| Anatomic study | Average internal Dice score (CI) | Average external Dice score (CI) | P value |
|---|---|---|---|
| Abdomen | 0.915 (0.902, 0.928) | 0.912 (0.898, 0.923) | 0.717 |
| Pelvis | 0.937 (0.930, 0.943) | 0.912 (0.906, 0.918) | < 0.001 |
| Prostate | 0.940 (0.922, 0.953) | 0.903 (0.896, 0.911) | < 0.01 |
| Shoulder | 0.890 (0.875, 0.903) | 0.904 (0.890, 0.915) | 0.990 |
| Knee | 0.937 (0.930, 0.943) | 0.927 (0.921, 0.933) | 0.466 |
| Cervical spine | 0.838 (0.825, 0.848) | 0.827 (0.817, 0.838) | 0.415 |
| Thoracic spine | 0.831 (0.819, 0.844) | 0.833 (0.817, 0.847) | 0.838 |
| Lumbar spine | 0.869 (0.856, 0.881) | 0.872 (0.858, 0.886) | 0.890 |
| **Overall average** | **0.878 (0.873, 0.884)** | **0.875 (0.869, 0.880)** | **0.899** |

Scores by anatomic study are averaged across the contained anatomic structures, while the overall average is across all 44 structures for the internal dataset and external dataset. 95% confidence intervals (CI) are reported in parentheses. P values from non-inferiority tests are provided; a p value < 0.05 rejects the null hypothesis that MRAnnotator demonstrates non-inferior performance on external datasets compared to the internal dataset.

**Table 3.** Analysis of MRAnnotator performance by manufacturer and field strength on the internal and external test sets.

| | **Average internal Dice score** | **Average external Dice score** | **P value** |
|---|---|---|---|
| **Field Strength** | | | |
| 3 T | 0.878 | 0.892 | 0.984 |
| 1.5 T | 0.878 | 0.875 | 0.889 |
| 0.55 T | 0.821 | n/a | n/a |
| 0.3 T* | n/a | 0.805 | n/a |
| **Manufacturer** | | | |
| GE | 0.865 | 0.878 | 1.0 |
| Philips | 0.865 | 0.934** | n/a |
| Siemens | 0.895 | 0.913** | n/a |
| Hitachi* | n/a | 0.805 | n/a |

*All 0.3 T studies were acquired using Hitachi open MRI machines, and all Hitachi studies were imaged at 0.3 T.

**Only 1 study.

**Table 4:** Benchmarking performance of MRAnnotator against a similar model trained on the AMOS abdominal MRI dataset, as well as two other public multi-anatomy MRI segmentation models

| | **MRAnnotator** | **AMOS nnU-Net** | **TSM** | **MRS** |
|---|---|---|---|---|
| Internal test set | 0.915 (0.902, 0.928) | 0.805 (0.780, 0.826; p<0.001) | 0.823 (0.807, 0.837; p<0.001) | 0.867 (0.850, 0.882; p<0.001) |
| External dataset | 0.912 (0.898, 0.923) | 0.879 (0.858, 0.897; p<0.001) | 0.822 (0.806, 0.837; p<0.001) | 0.869 (0.854, 0.882; p<0.001) |
| AMOS | 0.889 (0.866, 0.909) | 0.895 (0.871, 0.915; p=0.361) | 0.822 (0.800, 0.842; p<0.001) | 0.867 (0.844, 0.887; p<0.001) |

Parentheses include 95% CI and p values from paired t tests compared to the MRAnnotator results. Performance is evaluated on the 8 common abdominal classes from the AMOS test set and our internal and external datasets. TSM: TotalSegmentator MRI ; MRS: MRSegmentator

**Table 5:** Benchmarking multi-anatomy performance of MRAnnotator, TotalSegmentator MRI (TSM), and MRSegmentator (MRS) on the external dataset.

| **Anatomic Study (# Structures)** | **MRAnnotator** (Dice, 95% CI) | **TotalSegmentator MRI** (Dice, 95% CI) | **MRSegmentator** (Dice, 95% CI) |
|---|---|---|---|
| Abdomen (8) | 0.912 (0.898, 0.923) | 0.822 (0.806, 0.837) | 0.869 (0.854, 0.882) |
| Pelvis (3) | 0.912 (0.906, 0.918) | 0.848 (0.834,0.860) | 0.778 (0.770, 0.787) |
| Prostate (1) | 0.903 (0.895, 0.911) | 0.707 (0.675, 0.737) | n/a |
| Shoulder (3) | 0.925 (0.897, 0.947) | 0.335 (0.244, 0.426) | n/a |
| Knee (3) | 0.935 (0.928, 0.941) | 0.776 (0.750, 0.802) | n/a |

Performance was only evaluated on labeled structures shared by MRAnnotator, TSM, and MRS.

# SUPPLEMENTARY MATERIAL

S1: Model segmentation accuracy on internal and external dataset holdout test sets for all classes. P values are from non-inferiority tests, as described in the manuscript.

| Anatomic Study | # | Anatomic Structures | Internal Dice score (CI) | External Dice score (CI) | P values |
|---|---|---|---|---|---|
| Abdomen | 8 | Spleen | 0.950 (0.909, 0.973) | 0.959 (0.949, 0.966) | 0.773 |
| | | Right kidney | 0.961 (0.947, 0.972) | 0.970 (0.967, 0.973) | 0.979 |
| | | Left kidney | 0.969 (0.961, 0.975) | 0.972 (0.968, 0.975) | 0.996 |
| | | Gallbladder | 0.744 (0.669, 0.813) | 0.788 (0.713, 0.850) | 0.821 |
| | | Liver | 0.979 (0.972, 0.984) | 0.982 (0.981, 0.984) | 0.999 |
| | | Aorta | 0.963 (0.958, 0.968) | 0.888 (0.870, 0.906) | <0.001 |
| | | Inferior vena cava | 0.920 (0.907, 0.931) | 0.904 (0.893, 0.914) | 0.212 |
| | | Pancreas | 0.830 (0.791, 0.861) | 0.824 (0.788, 0.859) | 0.533 |
| Knee | 4 | Femur | 0.961 (0.955, 0.966) | 0.955 (0.949, 0.960) | 0.804 |
| | | Patella | 0.912 (0.895, 0.926) | 0.904 (0.892, 0.916) | 0.518 |
| | | Fibula | 0.911 (0.900, 0.921) | 0.894 (0.880, 0.908) | 0.201 |
| | | Tibia | 0.963 (0.957, 0.969) | 0.955 (0.949, 0.962) | 0.669 |
| Pelvis | 3 | Upper femur | 0.958 (0.949, 0.967) | 0.933 (0.922, 0.942) | <0.05 |
| | | Pelvis | 0.935 (0.924, 0.945) | 0.905 (0.895, 0.914) | <0.01 |
| | | Sacrum | 0.917 (0.904, 0.93) | 0.896 (0.886, 0.905) | 0.090 |
| Shoulder | 6 | Right humerus | 0.947 (0.931, 0.961) | 0.940 (0.909, 0.964) | 0.568 |
| | | Right clavicle | 0.896 (0.871, 0.921) | 0.928 (0.912, 0.943) | 0.996 |
| | | Right scapula | 0.861 (0.835, 0.885) | 0.886 (0.870, 0.90) | 0.987 |
| | | Left humerus | 0.942 (0.925, 0.957) | 0.909 (0.860, 0.947) | 0.161 |
| | | Left clavicle | 0.88 (0.838, 0.915) | 0.896 (0.861, 0.923) | 0.818 |
| | | Left scapula | 0.829 (0.79, 0.860) | 0.863 (0.832, 0.887) | 0.964 |
| C-Spine | 5 | Vertebrae C3-C7 | 0.841 (0.821, 0.86) | 0.823 (0.801, 0.846) | 0.263 |
| | | | 0.832 (0.809, 0.854) | 0.818 (0.793, 0.841) | 0.366 |
| | | | 0.83 (0.806, 0.852) | 0.809 (0.784, 0.832) | 0.226 |
| | | | 0.834 (0.802, 0.861) | 0.838 (0.814, 0.859) | 0.730 |
| | | | 0.851 (0.814, 0.876) | 0.850 (0.833, 0.867) | 0.649 |
| T-Spine | 12 | Vertebrae T1-T12 | 0.8 (0.76, 0.833) | 0.789 (0.747, 0.827) | 0.459 |
| | | | 0.84 (0.812, 0.865) | 0.827 (0.780, 0.867) | 0.421 |
| | | | 0.836 (0.807, 0.862) | 0.826 (0.777, 0.861) | 0.479 |
| | | | 0.832 (0.784, 0.87) | 0.821 (0.782, 0.855) | 0.468 |
| | | | 0.821 (0.774, 0.86) | 0.845 (0.798, 0.882) | 0.833 |
| | | | 0.825 (0.772, 0.862) | 0.838 (0.784, 0.882) | 0.725 |
| | | | 0.857 (0.823, 0.883) | 0.825 (0.758, 0.874) | 0.241 |
| | | | 0.848 (0.817, 0.876) | 0.837 (0.774, 0.886) | 0.480 |
| | | | 0.842 (0.801, 0.875)<br>0.842 (0.795, 0.881)<br>0.83 (0.775, 0.877)<br>0.806 (0.74, 0.859) | 0.826 (0.753, 0.886)<br>0.850 (0.796, 0.896)<br>0.859 (0.808, 0.902)<br>0.847 (0.791, 0.891) | 0.425<br>0.686<br>0.842<br>0.870 |
| L-Spine | 5 | Vertebrae L1-L5 | 0.857 (0.817, 0.891)<br>0.849 (0.812, 0.884)<br>0.878 (0.857, 0.899)<br>0.876 (0.856, 0.895)<br>0.885 (0.867, 0.9) | 0.855 (0.823, 0.883)<br>0.864 (0.837, 0.889)<br>0.880 (0.843, 0.908)<br>0.887 (0.851, 0.912)<br>0.875 (0.843, 0.898) | 0.599<br>0.848<br>0.696<br>0.826<br>0.488 |
| Prostate | 1 | Prostate | 0.940 (0.919, 0.952) | 0.903 (0.896, 0.911) | <0.01 |

S2: Combined model vs. sub-model performance by class on the internal test dataset.

| **Anatomic study** | **Sub-model internal Dice score** | **Combined-model internal Dice score** |
|---|---|---|
| Abdomen | 0.915 (0.902, 0.928) | 0.899 (0.885, 0.913) |
| Pelvis | 0.937 (0.930, 0.943) | 0.930 (0.923, 0.936) |
| Prostate | 0.940 (0.922, 0.953) | 0.941 (0.932, 0.948) |
| Shoulder | 0.890 (0.875, 0.903) | 0.880 (0.863, 0.897) |
| Knee | 0.937 (0.930, 0.943) | 0.934 (0.927, 0.940) |
| Cervical spine | 0.838 (0.825, 0.848) | 0.834 (0.822, 0.845) |
| Thoracic spine | 0.831 (0.819, 0.844) | 0.779 (0.763, 0.794) |
| Lumbar spine | 0.869 (0.856, 0.881) | 0.809 (0.780, 0.836) |
| **Overall average** | **0.878 (0.873, 0.884)** | **0.851 (0.844, 0.858)** |

S3: Performance on each shared anatomic structure class of the external dataset for MRAnnotator, TotalSegmentator MRI, and MRSegmentator.

| **Anatomic Study** | **#** | **Anatomic Structures** | **MRAnnotator (CI)** | **TotalSegmentator MRI (CI)** | **MRSegmentator (CI)** |
|---|---|---|---|---|---|
| Abdomen | 8 | Spleen<br>Right kidney<br>Left kidney | 0.959 (0.949, 0.966)<br>0.970 (0.967, 0.973)<br>0.972 (0.968, 0.975) | 0.886 (0.874, 0.898)<br>0.907 (0.896, 0.919)<br>0.894 (0.882, 0.906) | 0.929 (0.918, 0.939)<br>0.943 (0.935, 0.951)<br>0.944 (0.936, 0.952) |
| | | Gallbladder<br>Liver<br>Aorta<br>Inferior vena cava<br>Pancreas | 0.788 (0.713, 0.850)<br>0.982 (0.981, 0.984)<br>0.888 (0.870, 0.906)<br>0.904 (0.893, 0.914)<br>0.824 (0.788, 0.859) | 0.729 (0.666, 0.788)<br>0.941 (0.938, 0.945)<br>0.704 (0.662, 0.745)<br>0.774 (0.745, 0.799)<br>0.737 (0.705, 0.767) | 0.757 (0.702, 0.805)<br>0.967 (0.965, 0.970)<br>0.844 (0.812, 0.870)<br>0.809 (0.789, 0.827)<br>0.752 (0.714, 0.788) |
| Knee | 3 | Femur<br>Fibula<br>Tibia | 0.955 (0.949, 0.960)<br>0.894 (0.880, 0.908)<br>0.955 (0.949, 0.962) | 0.726 (0.654, 0.795)<br>0.722 (0.704, 0.740)<br>0.881 (0.875, 0.887) | - |
| Pelvis | 3 | Upper femur<br>Pelvis<br>Sacrum | 0.933 (0.922, 0.942)<br>0.905 (0.895, 0.914)<br>0.896 (0.886, 0.905) | 0.876 (0.860, 0.891)<br>0.874 (0.862, 0.885)<br>0.793 (0.766, 0.815) | 0.806 (0.795, 0.817)<br>0.773 (0.758, 0.786)<br>0.756 (0.742, 0.769) |
| Shoulder | 2 | Right humerus<br>Left humerus | 0.940 (0.909, 0.964)<br>0.909 (0.860, 0.947) | 0.419 (0.297, 0.552)<br>0.251 (0.127, 0.360) | - |
| Prostate | 1 | Prostate | 0.903 (0.896, 0.911) | 0.707 (0.669, 0.737) | - |